\documentclass[pra,twocolumn,superscriptaddress,showpacs,amsmath,amssymb]{revtex4-1}
\usepackage{graphicx}
\usepackage{mathrsfs}
\usepackage{bm}
\usepackage{amssymb}
\usepackage{amsmath}
\usepackage{dcolumn}
\usepackage{color}

\def\u#1{_{\rm #1}}
\newcommand{\ket}[1]{| #1 \rangle}

\newcommand{\ketbra}[2]{| #1 \rangle \langle #2 |}
\newcommand{\expect}[1]{\langle #1 \rangle} 
\def\H{{\rm H}}
\def\V{{\rm V}}

\def\00{\H\V}
\def\11{\V\H}

\begin{document}
\title{
Optimal conditions for Bell test using spontaneous
parametric down-conversion sources
}

\author{Yoshiaki~Tsujimoto}
\affiliation{Advanced ICT Research Institute, National Institute of Information and Communications
Technology (NICT), Koganei, Tokyo 184-8795, Japan\\}
\author{Kentaro~Wakui}
\affiliation{Advanced ICT Research Institute, National Institute of Information and Communications
Technology (NICT), Koganei, Tokyo 184-8795, Japan\\}
\author{Mikio~Fujiwara}
\affiliation{Advanced ICT Research Institute, National Institute of Information and Communications
Technology (NICT), Koganei, Tokyo 184-8795, Japan\\}
\author{Kazuhiro~Hayasaka}
\affiliation{Advanced ICT Research Institute, National Institute of Information and Communications
Technology (NICT), Koganei, Tokyo 184-8795, Japan\\}
\author{Shigehito Miki}
\affiliation{Advanced ICT Research Institute, National Institute of Information and Communications Technology~(NICT), 588-2 Iwaoka, Iwaoka-cho, Nishi-ku, Kobe, 651-2492, Japan\\}
\affiliation{Graduate School of Engineering Faculty of Engineering, Kobe University, 1-1 Rokko-dai cho, Nada-ku, Kobe 657-0013, Japan\\}
\author{Hirotaka~Terai}
\affiliation{Advanced ICT Research Institute, National Institute of Information and Communications Technology~(NICT), 588-2 Iwaoka, Iwaoka-cho, Nishi-ku, Kobe, 651-2492, Japan\\}
\author{Masahide~Sasaki}
\affiliation{Advanced ICT Research Institute, National Institute of Information and Communications
Technology (NICT), Koganei, Tokyo 184-8795, Japan\\}
\author{Masahiro~Takeoka}
\affiliation{Advanced ICT Research Institute, National Institute of Information and Communications
Technology (NICT), Koganei, Tokyo 184-8795, Japan\\}

\begin{abstract}
We theoretically and experimentally investigate the optimal conditions for 
the Bell experiment using spontaneous parametric down conversion~(SPDC) sources. 
In theory, we show that relatively large average photon number~(typically $\sim$0.5) is desirable to observe the 
maximum violation of the Clauser-Horne-Shimony-Holt~(CHSH) inequality. 
In experiment, 
we perform the Bell experiment without postselection 
using polarization entangled photon pairs at 1550~nm telecommunication wavelength generated from SPDC sources. 
While the violation of the CHSH inequality is not directly observed due to the overall detection efficiencies of our system, 
the experimental values agree well with those obtained by the theory with experimental imperfections. 
Furthermore, in the range of the small average photon numbers~($\leq0.1$), 
we propose and demonstrate a method to estimate the ideal CHSH value intrinsically contained in the tested state from the lossy experimental data 
without assuming the input quantum state.
\end{abstract}
\pacs{03.67.Hk, 03.67.Bg, 42.65.Lm}

\maketitle

\section{Introduction}

Quantum mechanically entangled photon pairs are essential tools for various optical quantum information and communication 
protocols~\cite{RevModPhys.79.135,RevModPhys.84.777}. 
Such entangled photon pairs can be generated with spontaneous parametric down conversion~(SPDC). 
To generate perfectly correlated pairs via the SPDC process which is probabilistic, it is frequently driven 
by weak pumping regime, such that emitted light only contains biphotons (a pair of single photons) and 
higher-order multi-photon emissions are sufficiently low. 
This feature is useful when one makes postselection of the coincidence photon counting events. 


The weakly-pumped SPDC source has also been used in the experiment {\it without} postselection. 
One important example is a loophole-free test of the Bell inequality~\cite{bell1964js,RevModPhys.86.419}. 
Violation of the Bell inequality rules out the possibility of describing the correlation between two parties by the local hidden variable model. 
To observe the genuine quantum correlation directly, it is important that the Bell test is performed without any loopholes, e.g. the detection loophole. 
In addition, the loophole-free Bell test implies new quantum information applications such as device-independent quantum key distribution~(DIQKD)~\cite{PhysRevLett.98.230501,pironio2009device} and 
random number generation~\cite{colbeck2012free}. 
So far, in photonic systems, 
the violation of the Bell inequality closing the detection and locality loopholes~\cite{giustina2013bell,PhysRevLett.111.130406,PhysRevLett.115.250401,PhysRevLett.115.250402} have been demonstrated by 
combining the weakly-pumped SPDC sources and highly efficient detectors.

Though these experiments successfully violates the Clauser-Horne-Shimony-Holt~(CHSH) 
inequality~\cite{PhysRevLett.23.880}, the amount of violation was limited to be small~($\sim10^{-4}$)  
since the weakly-pumped SPDC source 
mainly emits vacuum and only a few biphotons. 
The average photon number is typically in the order of $10^{-2}$.
That is, the major component of the quantum state is vacuum, which does not contribute to yield 
the violation of the CHSH inequality. 
In contrast, very recently, larger violation of the CHSH inequality have been reported 
by using strongly-pumped SPDC sources 
which produce a non-negligible amount of multiple pairs~\cite{PhysRevLett.120.010503,Lijiong2018}. 
Moreover, the theoretical analysis~\cite{PhysRevA.91.012107} considering multi-photon pair
emissions of the SPDC sources  
indicates that the maximum violation of the CHSH 
inequality is $\sim0.35$ which is much larger than those obtained in the previous experiments~\cite{giustina2013bell,PhysRevLett.111.130406,PhysRevLett.115.250401,PhysRevLett.115.250402}. 
Thus, further study is required for clarifying the best quantum state which maximizes the CHSH inequality violation.


In this paper, both theoretically and experimentally, we elucidate the optimal conditions for SPDC sources to achieve the maximum violation of the CHSH inequality. 
First we construct a realistic model based on the characteristic function approach, which can take into account higher-order multi-photon pair 
emissions~\cite{takeoka2015full,PhysRevA.93.042328}. 
Then we show the optimal parameters for the 
system for a given detection efficiency~($\eta$) in detail, 
including average photon numbers~($\lambda$) of the two SPDC sources and their relative ratio, and optimal measurement angles.
It is revealed that the maximal violation is obtained at relatively high average photon number regime 
where the contribution of multi-photon pair emissions is not negligible: 
$\lambda>0.1$ in most cases, and $\lambda=0.99$ is optimal for $\eta=1$. 
We also show that the measurement angle of the Bell test is almost independent of the detection efficiency. 
It is noteworthy that this feature allows us to reduce the number of optimization parameters, and therefore is practically useful for saving computational resources.

Second, to test the theoretical predictions, we perform the Bell-test experiment 
without postselections
using polarization entangled photon pairs generated by SPDC. 
We collected all the events including no-detection~(vacuum) events, and 
calculated the CHSH value for each average photon number.  
While the overall detection efficiencies of our system are 
insufficient to directly observe the violation of the CHSH inequality, 
the CHSH values obtained by the experiment 
well agree with the theory in a wide range of parameters. 
Furthermore, for the low average photon number regime of $\lambda\leq0.1$, 
we propose and demonstrate a method to 
estimate the ideal probability distributions of the Bell test  
from the lossy experimental data without assuming the input quantum state.  
The results agree with the theory and thus provide a useful estimation technique 
for quantum optics experiments with certain amount of losses. 

The paper is organized as follows. In Sec.~\ref{secII}, we briefly review the Bell test using SPDC sources 
and describe our theoretical model 
including higher order photon numbers and experimental imperfections. 
In Sec.~\ref{secIII}, we present our numerical results. 
The experimental setup is described in Sec.~\ref{secIV}. 
In Sec.~\ref{secV}, we present our experimental results and 
introduce the method to compensate the loss of the system. 
We conclude the paper in Sec.~\ref{secVI}.

\section{Bell test via the SPDC sources}
\label{secII}
 \begin{figure}[t]
 \begin{center}
 \includegraphics[width=\columnwidth]{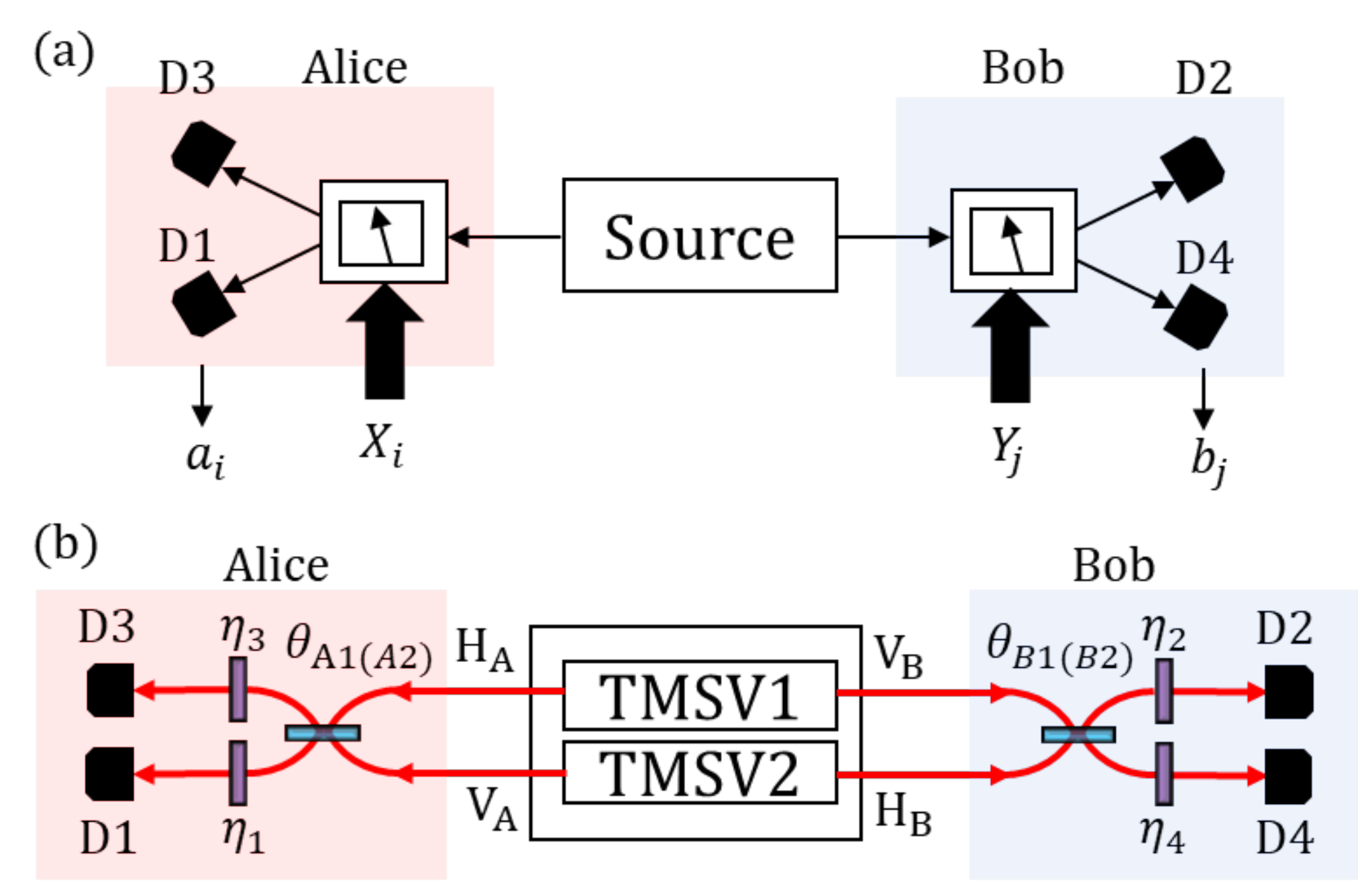}
  \caption{(a)~The schematic diagram for the Bell experiment. Alice and Bob share a pair of 
	particles, and choose the measurement settings $X_i\in\{X_1, X_2\}$ 
	and $Y_j\in\{Y_1, Y_2\}$, respectively. The measurement outcomes are 
	binary, i.e., $a_i, b_j\in\{-1, +1\}$. (b)~The realistic model for the Bell experiment. 
	An entangled photon pair is generated by means of a pair of two-mode squeezed 
	vacua~(TMSV) over polarization modes. The polarization measurement is realized 
	by the polarization mixing followed by the on-off type, single photon detectors with dark counts.
     \label{fig:BellSetup}}
 \end{center}
\end{figure}
The schematic diagram of the Bell test is shown 
in Fig.~\ref{fig:BellSetup}(a). A pair of particles is distributed from the source to two receivers, Alice and Bob. 
They randomly choose the measurement settings $X_i\in\{X_1, X_2\}$ and 
$Y_j\in\{Y_1, Y_2\}$, respectively. All the observables produce 
binary outcomes $a_i, b_j\in\{-1, +1\}$. Alice and Bob repeat the measurement, 
and calculate the CHSH value 
\begin{equation}
S=\expect{a_1b_1}+\expect{a_2b_1}+\expect{a_1b_2}-\expect{a_2b_2},
\label{eq:CHSH}
\end{equation}
where $\expect{a_ib_j}=P(a=b|X_i,Y_j)-P(a\neq b|X_i,Y_j)$. 
Here, $S>2$ indicates that the particles shared between Alice and Bob possess 
nonlocal quantum correlation which cannot be reproduced by any local hidden variables. 
The maximum value of $S$ allowed by quantum mechanics is $2\sqrt{2}$, 
which is known as the Cirelson bound~\cite{cirel1980bs} and 
achieved by using a maximally entangled pair. 

Next, the realistic model of the Bell test with SPDCs is shown in Fig.~\ref{fig:BellSetup}(b). 
The SPDCs emit entangled photon pairs, or more precisely, the two-mode squeezed vacuum~(TMSV) whose 
Hamiltonian is represented by $\hat{H}=i\hbar(\zeta_1\hat{a}^\dagger_{H_A}\hat{a}^\dagger_{V_B}+\zeta_2\hat{a}^\dagger_{V_A}\hat{a}^\dagger_{H_B}-\rm{h.c.})$,
where $\hat{a}_{j}^\dagger$ is the photon creation operator in mode $j$, 
and $\zeta_k=|\zeta_k|e^{i\phi_k}$ is the coupling constant 
of TMSV$k$ ($k=1,2$) which is proportional to the complex amplitude of each pump. 
In the following, $\phi_k$ is fixed as $\phi_1=0$ and $\phi_2=\pi$. 
$H$ and $V$ denote the horizontal and vertical polarizations, respectively. 
The generated quantum state is described by 
\begin{eqnarray}
\ket{\Psi_{\mathrm{ent}}}&=&\mathrm{exp}(-i\hat{H}t/\hbar)\ket{0}\\
&=&\sum^{\infty}_{n=0}\frac{1}{\mathrm{cosh}r_1\mathrm{cosh}r_2}\sqrt{n+1}\ket{\Phi_n},
\end{eqnarray}
where 
\begin{equation}
\ket{\Phi_n}=\frac{(-i)^n}{n!\sqrt{n+1}}(\mathrm{tanh}r_1\hat{a}^\dagger_{H_A}\hat{a}^\dagger_{V_B}-\mathrm{tanh}r_2\hat{a}^\dagger_{V_A}\hat{a}^\dagger_{H_B})^n\ket{0}.
\label{output3}
\end{equation}
Here $\ket{0}$ is the vacuum state, and $r_k=|\zeta_k|t$ is the squeezing parameter of TMSV$k$. 
Note that the average photon number of TMSV$k$ is given by $\lambda_k=\mathrm{sinh}^2r_k$. 
The state clearly consists of an infinite series and the contribution from higher order photon numbers cannot be negligible even with finite $\lambda_k$.
The polarizer with angle $\theta$ works as a polarization-domain beamsplitter mixing the $H$ and $V$ modes where  
its transmittance and reflectance are $\mathrm{cos}^2\theta$ and $\mathrm{sin}^2\theta$, respectively. 
The overall detection efficiencies including 
the system transmittance and the imperfect quantum efficiencies of the detectors 
are denoted by $\eta_l$ for $l=1, 2, 3, 4$.
This is modeled by inserting the losses in each arm before the detectors with unit efficiency 
(see Fig.~\ref{fig:BellSetup}(b)). 
We consider that the detectors D1, D2, D3 and D4 are on-off type, single photon detectors with dark counts 
which only distinguish between vacuum (off: no-click) and non-vacuum (on: click)
with dark count probability of $\nu$. 
\section{Numerical results}
\label{secIII}

 \begin{figure}[t]
 \begin{center}
 \includegraphics[width=\columnwidth]{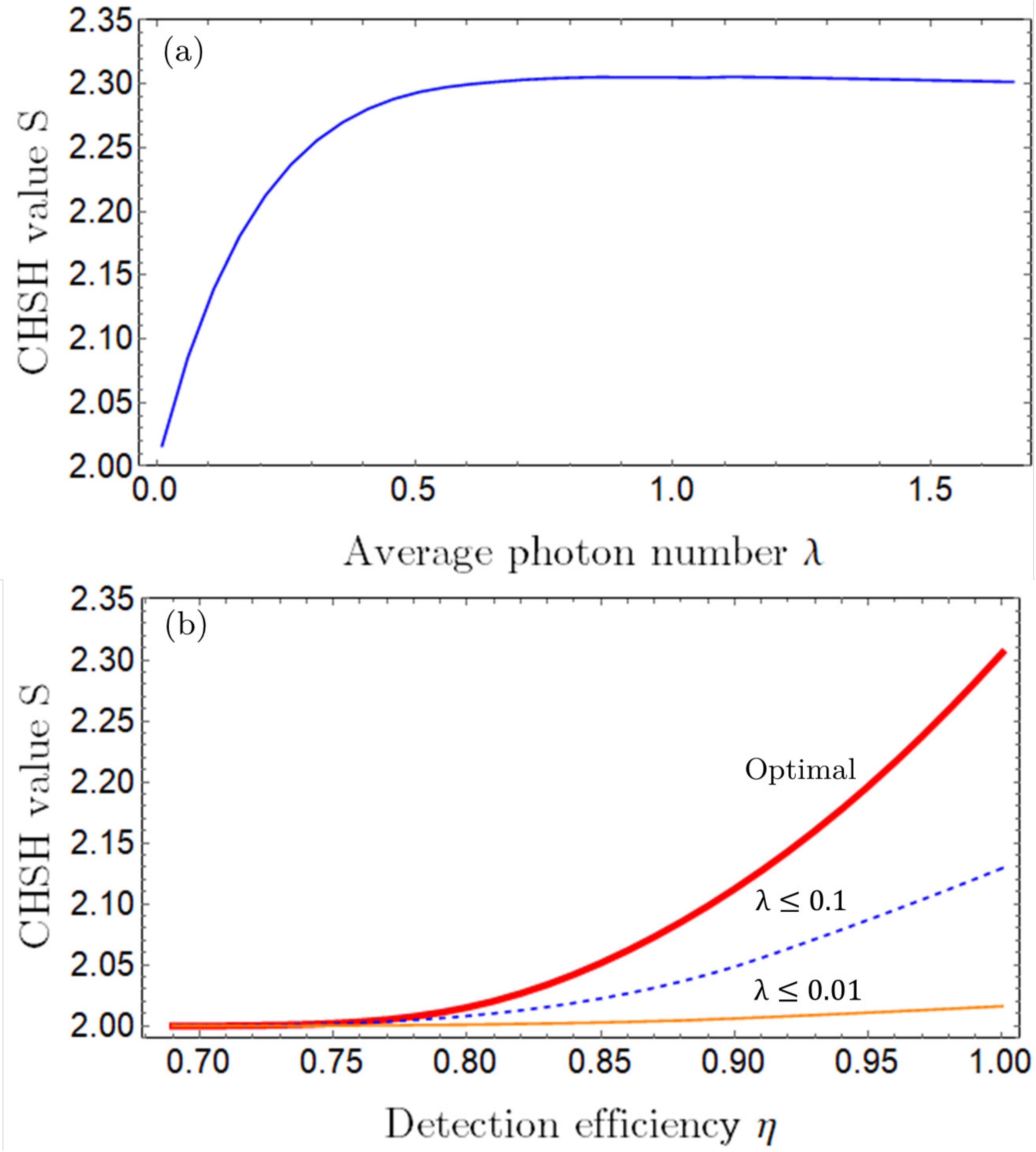}
  \caption{(a)~The average photon number $\lambda$ vs $S$. 
	For each point, we fix $\lambda$, and optimize the other parameters. 
	 (b)~The overall detection efficiency $\eta$ vs $S$ for the three different ranges of $\lambda$. 
	For each $\eta$, we fix the ranges of $\lambda$, and perform optimizations.}	
\label{fig:theory}
 \end{center}
\end{figure}

\begin{figure}[t]
 \begin{center}
 \includegraphics[width=\columnwidth]{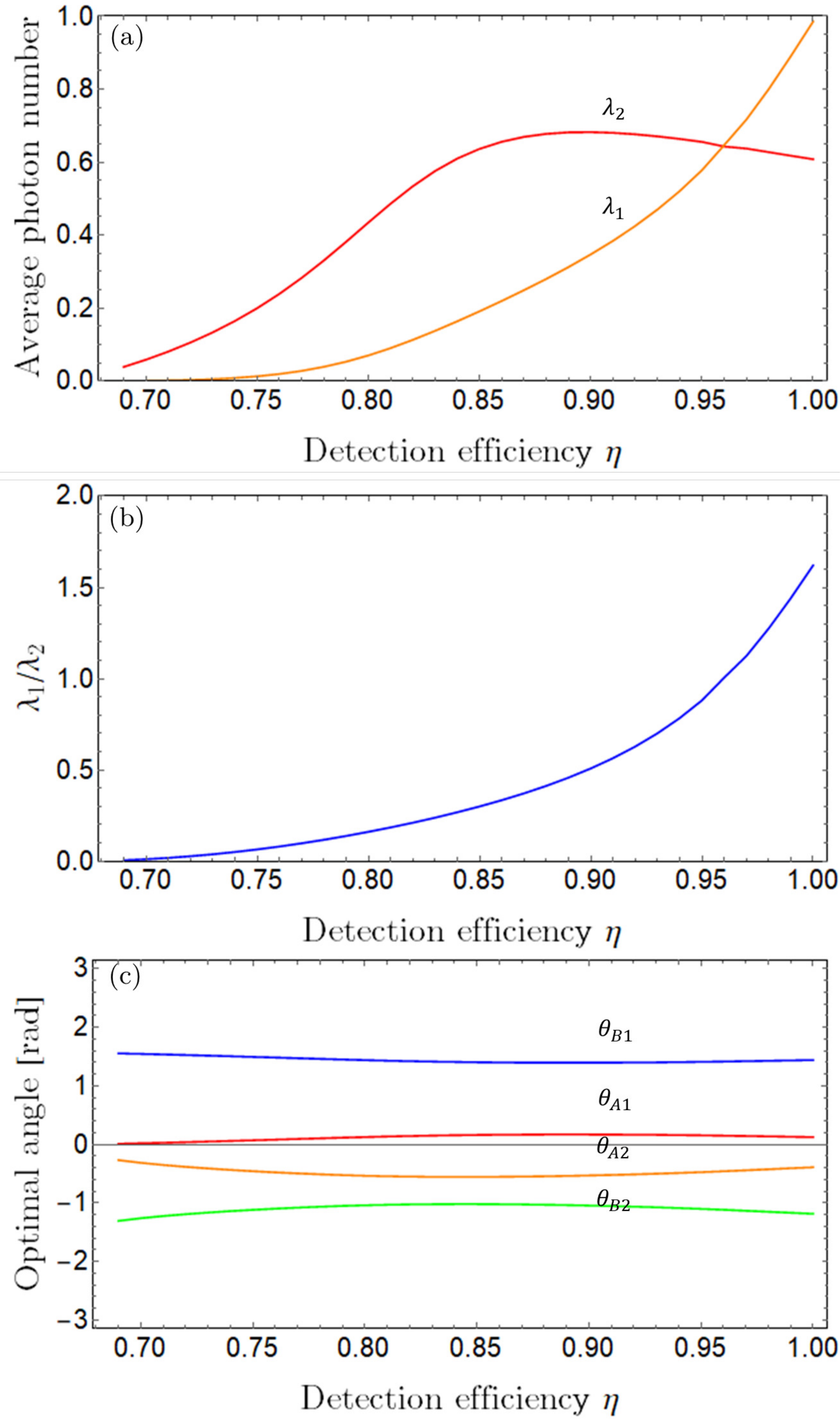}
  \caption{(a)~The overall detection efficiency $\eta$ vs the optimal average photon numbers~($\lambda_1$ and $\lambda_2$). 
		   (b)~$\eta$ vs $\lambda_1/\lambda_2$. 
	 (c)~$\eta$ vs the optimal angles of the polarizers. In the simulations, we assume that $\eta_1=\eta_2=\eta_3=\eta_4:=\eta$ and $\nu=0$.}	
\label{fig:theoryA}
 \end{center}
\end{figure}

To numerically calculate $S$ in Eq.~(\ref{eq:CHSH}) with the above SPDC model, we use the approach based on 
the characteristic function~\cite{takeoka2015full,PhysRevA.93.042328}. 
This approach is applicable when the system is composed of Gaussian states and operations, and on-off detectors. 
The Gaussian state is defined by the state whose characteristic function (or equivalently Wigner function) has a Gaussian distribution, 
including TMSV states. 
The Gaussian operation is also defined as an operation transforming a Gaussian state to another Gaussian state, 
which includes the operations by linear optics and second-order nonlinear processes. 
The setup in Fig.~\ref{fig:BellSetup}(b) includes only these means and thus meets the condition above. 
See Appendix and Ref.~\cite{takeoka2015full} for more details of this approach. 
Note that a similar calculation with a different approach is reported in Ref.~\cite{PhysRevA.91.012107}. 

We calculate the probability of all the combinations of the photon detection (click) and no-detection (no-click) events
for each polarizer angle, and obtain the probability distributions. 
We denote, for example, the probability of observing clicks in D1 and D2, and no-clicks 
in D3 an D4 as $P(\mathrm{c1,c2,nc3,nc4})$. 
Each of Alice and Bob determines her/his local rule, and 
assign +1 or $-1$ for each detection event. 
Since there are four possible local events for each of Alice and Bob, i.e., (i)~only the one detector clicks, (ii)~only the other detector clicks, (iii)~both of the two detectors simultaneously click and (iv)~no detector clicks, 
there are 16 possible choices for each of Alice and Bob to assign $\pm1$.
We introduce the following simple local assignment strategy for Alice~(Bob): only D1(D2) clicks$\rightarrow-1$ and otherwise $\rightarrow+1$, respectively. 
Under the condition that Alice~(Bob) chooses the angle $\theta_{A1}~(\theta_{B1})$, respectively, the 
probability that both of Alice and Bob obtain the outcome $-1$ 
is calculated by 
$P(-1,-1|\theta_{A1},\theta_{B1})$=$P(\mathrm{c1,c2,nc3,nc4})$. 
Similarly, the other conditional probabilities $P(+1,-1|\theta_{A1},\theta_{B1})$, $P(-1,+1|\theta_{A1},\theta_{B1})$ and $P(+1,+1|\theta_{A1},\theta_{B1})$ are also calculated by the detection probabilities, 
which enables us to calculate $S$. 
See Appendix for the details of the formulas.

 \begin{figure*}[t]
 \begin{center}
\scalebox{0.35}{\includegraphics{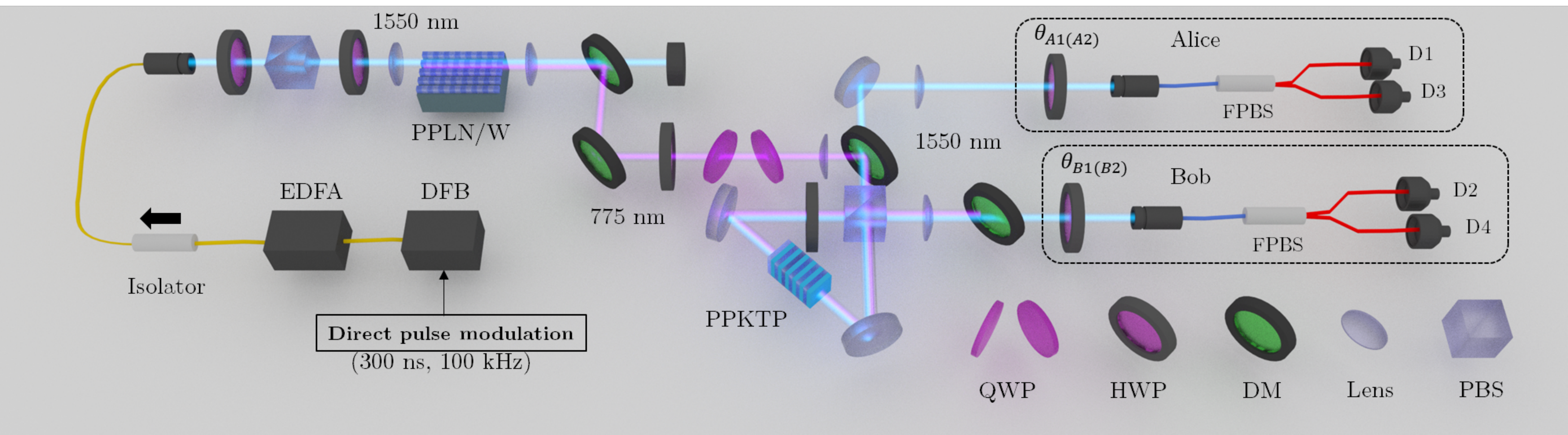}}
  \caption{The setup for the Bell experiment. To generate entangled photon pairs by SPDC, 
we used counter propagating pump pulses to excite the PPKTP crystal in the Sagnac loop interferometer. 
Alice and Bob choose the measurement angles 
$\{\theta_{A1},\theta_{A2}\}$ and $\{\theta_{B1},\theta_{B2}\}$, respectively, 
and assign +1 or -1 for the each detection event to
calculate $S$ value. DFB: distributed feedback laser, EDFA: erbium-doped fiber amplifier, 
PPLN/W: periodically poled lithium niobate waveguide, PPKTP: periodically poled potassium titanyl phosphate, 
QWP: quarter waveplate, HWP: half waveplate, DM: dichroic mirror, PBS: polarization beamsplitter, FPBS: 
fiber-based PBS.}
\label{fig:experiment}
 \end{center}
\end{figure*}


Fig.~\ref{fig:theory}(a) shows the relation between the average photon number and $S$ 
in an ideal system, where all the detection efficiencies are unity and detectors have no dark counts
(i.e. $\eta_1=\eta_2=\eta_3=\eta_4=1$ and $\nu=0$).
We define 
the larger of $\lambda_1$ and $\lambda_2$ as $\lambda$, 
and then for given $\lambda$, numerically optimize the other average photon number 
and $\{\theta_{A1(B1)}, \theta_{A2(B2)}\}$ via the Nelder-Mead method such that $S$ is maximized. 
$S$ at the maximum violation is around 2.31, 
which coincides with the theoretical result 
by Vivoli {\it et al}.~\cite{PhysRevA.91.012107}.
We found that 
the maximum violation is obtained at $\lambda=0.99$ which is much larger than those used in the previous experiments~\cite{giustina2013bell,PhysRevLett.111.130406,PhysRevLett.115.250401,PhysRevLett.115.250402}. 
Note that the maximum $S$ obtained in Fig.~\ref{fig:theory}(a) is robust against the dark counts. 
In fact, when we add a dark count probability of $\nu=10^{-4}$ to each detector as a realistic value, degradation of $S$ was as small as $5.0\times10^{-4}$.

Next, we show the loss tolerance of $S$ for the three different ranges of $\lambda$ in Fig.~\ref{fig:theory}(b).
In the simulation, we assumed that $\eta_1=\eta_2=\eta_3=\eta_4:=\eta$ and $\nu=0$. 
$\lambda_1$, $\lambda_2$ and the measurement angles are optimized for each $\eta$. 
The red thick curve, blue dashed curve, and yellow thin curve represent 
$S$ optimized under the conditions of $0\leq\lambda$, $0\leq\lambda\leq0.1$, and $0\leq\lambda\leq0.01$, respectively. 
The figure shows that the limited $\lambda$ strongly restricts the maximum $S$ in any $\eta$. 
The result indicates that the maximum $S$ obtainable in the previous Bell experiments 
using SPDC sources with small $\lambda$ is intrinsically limited and thus 
suggests a use of higher pumping of the SPDC sources to obtain larger CHSH violation.  

Finally, we show the optimal parameters for given $\eta$ in Figs.~\ref{fig:theoryA}(a)-(c). 
The optimal average photon numbers are shown in Fig.~\ref{fig:theoryA}(a). 
Even when $\eta=1$, the two optimal average photon numbers are unbalanced. 
The ratio between $\lambda_1$ and $\lambda_2$ is shown in Fig.~\ref{fig:theoryA}(b). 
We found that the ratio of $\lambda_1/\lambda_2$ monotonically and continuously decreases as 
$\eta$ decreases. 
This result qualitatively agrees with the analysis based on qubit systems in Ref.~\cite{PhysRevA.47.R747}. 
The optimal angles of the polarizers are shown in Fig.~\ref{fig:theoryA}(c). 
Interestingly, the optimal angles are almost constant regardless of $\eta$. 


 \section{Experimental setup}
\label{secIV}
Theoretical predictions in the above section are 
verified using the experimental setup illustrated in Fig.~\ref{fig:experiment}.
We choose the measurement angles as 
$\{\theta_{A1},\theta_{A2}\}=\{0, \pi/5\}$ and $\{\theta_{B1},\theta_{B2}\}=\{3\pi/5, -3\pi/5\}$ by which 
$S$ is expected to be $S=2.30$ with $\lambda_1=\lambda_2=0.62$ when the 
overall detection efficiency is unity and the dark count probabilities are zero. 
These angles are slightly different from those shown in Fig.~\ref{fig:theoryA}(b) 
since we apply the condition $\lambda_1=\lambda_2$ for simplicity. 
A distributed feedback~(DFB) laser generates pulsed light at 1550~nm. 
The DFB laser is directly modulated by electrical pulses with 100~kHz repetition and 300~ns duration.
The output laser pulse is amplified by an erbium-doped fiber amplifier~(EDFA). 
The output of EDFA is vertically polarized  by a half-waveplate~(HWP) and a polarizing beamsplitter~(PBS), and then coupled 
to the 34~mm-long type-0 periodically poled lithium niobate waveguide~(PPLN/W) for second harmonic generation~(SHG). 
Amplified spontaneous emission from the EDFA and unconverted fundamental light of the SHG 
are removed by the dichroic mirrors~(DMs). 
The polarization of the SHG pulses are adjusted by using 
a HWP and a pair of quarter waveplates~(QWPs).  
The maximum pulse energy~(average power) of our SHG pulses is 0.2~$\mu$J~(20~mW).
To generate polarization entangled photon pairs by SPDC process, SHG pulses are used to pump a 30~mm-long, type II, periodically poled  potassium titanyl phosphate~(PPKTP) crystal in a Sagnac loop interferometer with a PBS~\cite{Jin:14}.
The two-qubit component of the generated state forms a maximally entangled state 
$\ket{\Psi^-}=(\ket{HV}-\ket{VH})/\sqrt{2}$, where $\ket{H}$ and $\ket{V}$ denote the $H$ and $V$ polarization state of a single photon, respectively.  
One half of the photon pair passes through the DM and goes to Alice's side while the other photon 
goes to Bob's side.   
Alice and Bob set measurement angles $\{\theta_{A1},\theta_{A2}\}$ and $\{\theta_{B1},\theta_{B2}\}$, respectively,  
by means of the HWPs and fiber-based PBSs~(FPBSs). 
Finally, the photons are detected by four 
superconducting single photon detectors~(SSPDs) D1 and D3 for Alice, and D2 and D4 
for Bob, respectively. The quantum efficiencies of these SSPDs are around $75~\%$~\cite{Miki:17}. 
The dark count probabilities of the SSPDs are $3.0\times10^{-4}$ per a detection window of 300~ns corresponding to the pulse duration.
The modulation signal for the DFB laser is also used as a start signal for a time-to-digital converter (TDC), 
and the detection signals from D1, D2, D3 and D4 are used as stop signals of the TDC. 
All combination of click and no-click events are collected without postselection. 
We assign events of D1~(D2) clicks on Alice's~(Bob's) side as -1 and all the others as +1, then calculate $S$.

\section{Experimental results}
\label{secV}
Before performing the Bell-test experiment, we estimate the overall detection efficiencies $\eta_l$.  
Suppose a TMSV is detected by two detectors, D1 and D2. 
The overall detection efficiencies of the two modes ($\eta_1$ and $\eta_2$) are well estimated by following equation~\cite{klyshko1980use},
\begin{equation}
\eta_{1(2)}=\frac{C_{12}}{S_{2(1)}}.
\end{equation}
Here $C_{12}$ is the coincidence count between D1 and D2, and $S_{2(1)}$ is the single detection count at D2(1). 
Note that the average photon number of the TMSV photons is small enough for this measurement.
In our theoretical model shown in Fig.~\ref{fig:BellSetup}(b), 
we have assumed the same detection efficiencies for TMSV1 and TMSV2. 
Thus, in the experiment, we carefully align the optical system 
such that the overall detection efficiencies for TMSV1 and TMSV2 are the same as each other.  
We estimated them as 
$\eta_1=10.48\pm0.69~\%$, $\eta_2=12.76\pm0.97~\%$, $\eta_3=12.72\pm0.53~\%$ and $\eta_4=11.86\pm0.24~\%$. 


Once $\eta_l$ is estimated, the average photon number~($\lambda_k$) 
of TMSV$k$ is calculated by using following relation: 
 \begin{equation}
\frac{S_{1(2)}}{N}=\frac{\lambda_k\eta_{1(2)}}{1+\lambda_k\eta_{1(2)}}, 
\end{equation}
where 
$N$ is the number of the total events, which corresponds to the number of the start signals of the TDC. 

\begin{figure}[t]
 \begin{center}
 \includegraphics[width=\columnwidth]{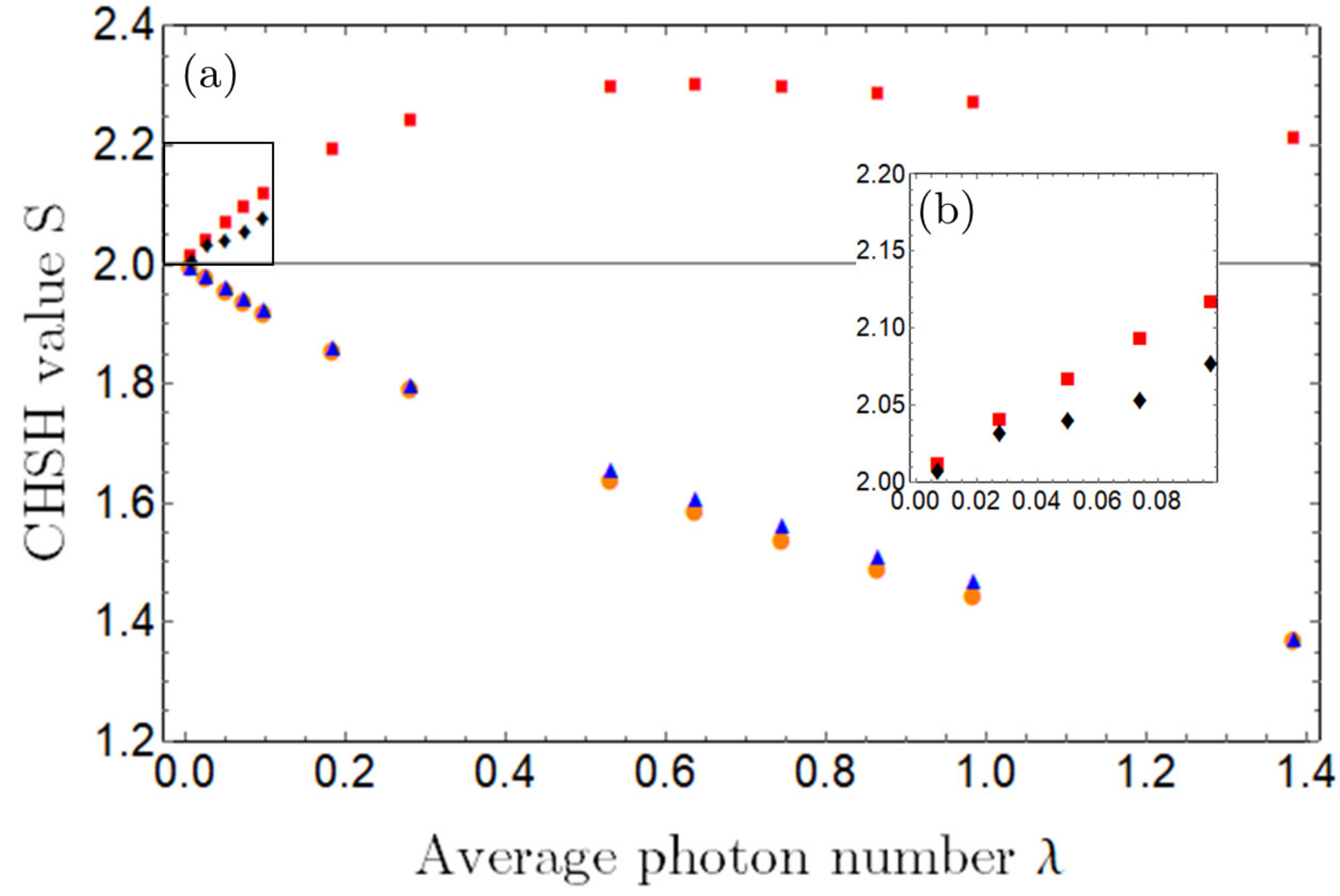}
  \caption{(a)~The $S$ values obtained by 
	the theory with unity detection efficiencies~(red square), 
	the theory with experimental parameters~(yellow circle) and 
	the experimental results~(blue triangle) for 
	the various values of $\lambda$. 
	The black diamonds represent the $S$ values obtained by compensating the losses of the system in the range of 
	 $\lambda\leq0.1$. (b)~The enlarged figure of the enclosed part.}	
\label{fig:result1}
 \end{center}
\end{figure}

In the Bell-test experiment, the difference between $\lambda_1$ and $\lambda_2$ is 
set to be less than 1~\%. 
Thus we denote $\lambda_1=\lambda_2:=\lambda$ in the following. 
The results of the Bell experiment is shown in Fig.~\ref{fig:result1}(a). 
We perform the Bell experiment for 
various values of $\lambda$ 
by changing the energy of the pump pulse. 
Though the overall detection efficiencies of our system are 
not in the range of directly observing the CHSH violation, 
it is still possible to compare our experimental results and the theory calculated 
with experimentally observed parameters: the average photon numbers, 
measurement angles, detection efficiencies and dark counts.
The experimental results~(blue triangles) and 
theoretical values with experimental parameters~(yellow circles) 
are in good agreement for each $\lambda$, which indicates that 
the theoretical model well explain the experimental results.  

In the low average photon number regime ($\lambda\leq0.1$), 
it is possible to compensate the imperfection of the overall detection efficiencies {\it without} 
assuming the quantum states distributed to Alice and Bob. 
In other words, one can estimate the intrinsic nonlocality
that could be observed with the unity detection efficiencies. 
Under the assumption that each detector detects at most one photon, 
the experimentally-obtained probability distribution 
$\bm{P}_{\mathrm{exp}}=(p_1, p_2,\cdots, p_{16})^T$ 
composed of the 16 combinations of the detection probabilities and 
the ideal probability distribution with the unity detection efficiencies 
$\bm{Q}_{\mathrm{ideal}}=(q_1, q_2,\cdots, q_{16})^T$ are 
connected by the linear transmission matrix $\bm{T}$ as 
\begin{equation}
\bm{P}\u{exp}=\bm{T}\bm{Q}\u{ideal}
\end{equation} 
for each measurement setting. 
Here, $p_1=P$(nc1,nc2,nc3,nc4), $p_2=P$(c1,nc2,nc3,nc4), and so on. 
$\bm{T}$ is the upper triangular matrix whose matrix elements are composed of the products of $\eta_l$ and $(1-\eta_l)$. 
For example, the four-fold coincidence probabilities $p_{16}$ and $q_{16}$ are connected by 
$p_{16}=q_{16}\prod_{i=1}^{4}\eta_i$. 
One may think that $\bm{Q}\u{ideal}$ is estimated by simply calculating 
$\bm{Q}\u{ideal}=\bm{T}^{-1}\bm{P}\u{\mathrm{exp}}$. 
However, in this case, the elements of $\bm{Q}\u{ideal}$ could be negative 
since $\bm{P}\u{exp}$ contains experimental errors. 
Thus we determine the most likely elements of $\bm{Q}\u{ideal}$ such that the $L^2$ distance between $\bm{P}\u{exp}$ and 
$\bm{T}\bm{Q}\u{ideal}$ is minimum under the condition that 
$q_i\geq0$ and $\sum_{i=1}^{16} q_i=1$. 
Namely, we estimate the probability distribution $\bm{Q}\u{ideal}$ which minimizes the function: 
\begin{equation}
\sum_{i=1}^{16}(p_i-\sum_{j=1}^{16}T_{ij}q_j)^2. 
\end{equation}
The $S$ values calculated by $\bm{Q}\u{ideal}$  
is shown in Fig.~\ref{fig:result1}(a) and (b) by the black triangles. 
The results agree with, but slightly below the theory plots for ideal state and detectors (red square), 
which reflects the deviation of the generated state from ideal TMSVs due to experimental imperfections. 
In particular, these two plots start to deviate in $\lambda>0.05$ where the probability of detecting 
multi-photon at each detector starts to be non-negligible.

\section{Conclusion}
\label{secVI}
In conclusion, we theoretically and experimentally investigate the optimal conditions for 
the Bell test with the SPDC sources. 
We construct a numerical model including multi-photon emissions from the SPDC sources and various imperfections, 
and see the maximum violation of the CHSH inequality as $S=2.31$ which agrees with the previous result 
in Ref.~\cite{PhysRevA.91.012107}.  
Then we show that the optimal experimental parameters to maximize 
the CHSH values for given average photon number 
of TMSV or the overall detection efficiency by numerical simulations. 
In particular, we show the CHSH value takes its maximum 
when the average photon number is much larger than those utilized in the previous experiments~\cite{giustina2013bell,PhysRevLett.111.130406,PhysRevLett.115.250401}. 
Next, we perform the Bell-test experiment without postselection using polarization entangled photon pairs 
generated by SPDC to test these theoretical predictions. 
The experimentally-obtained CHSH values agree well with those obtained by the theory. 
Moreover, in the range of small average photon numbers, 
we also propose and demonstrate a method to estimate the CHSH value 
of the quantum state before undergoing losses, 
by compensating the detection losses without assuming the input quantum state. 
The result shows good agreement with the theory model in the range of $\lambda\leq0.1$.
This approach is useful to estimate the property of quantum states 
via imperfect detectors. 

\begin{acknowledgments}
We thank Kaushik~P.~Seshadreesan for helpful discussions. 
This work was supported by JST CREST Grant No.~JPMJCR1772, 
MEXT/JSPS KAKENHI Grant No.~JP18K13487 and No.~JP17K14130. 
\end{acknowledgments}

\appendix*
\section{Detailed calculations based on the characteristic function}
We describe a procedure to calculate the 
probability distributions and the CHSH value $S$ using 
the theoretical model given in Sec.\ref{secII}. 
First, we review the basic tools used in the characteristic function approach  
which is often used in Gaussian continuous-variable quantum systems.  
This method allows us to deal with the quantum state generated 
by the SPDC process without the need for any approximations such as photon number truncation.  
Next, we present the method to calculate the detection probabilities.
Finally we describe the procedure to calculate $S$ 
using the obtained probability distribution. 

\subsection{Preliminary}
\subsubsection*{$\bold{Characteristic}$ $\bold{function}$}
Let us consider $N$ bosonic modes associated with a tensor product 
Hilbert space ${\mathcal H^{\otimes N}}=\bigotimes_{j=1}^N\mathcal{H}_j$, 
where $\mathcal{H}_j$ is an infinite dimensional Hilbert space. 
We define annihilation and creation operators corresponding each mode 
as $\hat{a}_j$ and $\hat{a}^\dagger_j$, respectively. They satisfy the commutation 
relation given by 
\begin{equation}
[\hat{a}_j,\hat{a}^\dagger_k]=\delta_{jk}.
\end{equation}
We also define the quadrature operators of a bosonic mode as 
\begin{eqnarray}
\hat{x}_j=\frac{1}{\sqrt{2}}(\hat{a}_j+\hat{a}^\dagger_j),\\
\hat{p}_j=\frac{1}{\sqrt{2}i}(\hat{a}_j-\hat{a}^\dagger_j). 
\end{eqnarray}
Note that we choose as a convention $\hbar=\omega=1$. 
Their commutation relation is calculated as 
\begin{equation}
[\hat{x}_j,\hat{p}_k]=i\delta_{jk}. 
\end{equation} 

We define a density operator acting on $\mathcal{H}^{\otimes N}$ as $\hat{\rho}$. 
The characteristic function of $\hat{\rho}$ is defined by 
\begin{equation}
\chi(\xi)=\mathrm{Tr}[\hat{\rho}\hat{\mathcal{W}}(\xi)],
\end{equation}
where 
\begin{equation}
\hat{\mathcal{W}}(\xi)=\mathrm{exp}(-i\xi^T\hat{R})
\end{equation}
is the Weyl operator. Here, 
$\hat{R}=(\hat{x}_1,\dots,\hat{x}_N,\hat{p}_1,\dots,\hat{p}_N)$ and 
$\xi=(\xi_1,\dots,\xi_{2N})$ 
are a 2$N$ vector consisting of quadrature operators and a 2$N$ real vector, respectively. 

\subsubsection*{$\bold{Gaussian}$ $\bold{states}$}
A Gaussian state is a quantum state whose characteristic function 
has a Gaussian distribution: 
\begin{equation}
\chi(\xi)=\mathrm{exp}(-\frac{1}{4}\xi^T\gamma\xi-id^T\xi), 
\end{equation}
where $\gamma$ is a 2$N$ $\times$ 2$N$ matrix called the covariance matrix and 
$d$ is a 2$N$-dimensional vector known as the displacement vector. 
The covariance matrix of the TMSV state generated by a SPDC source is given by 
\begin{equation}
\gamma^{\mathrm{TMSV}}(\lambda)=\left[
	\begin{array}{cc}
	\gamma^+(\lambda)&\bold{0}\\
	\bold{0}&\gamma^-(\lambda)
	\end{array}
	\right], 
\end{equation}
where
\begin{equation}
\gamma^{\mathrm{\pm}}=\left[
	\begin{array}{cc}
	2\lambda+1&\pm2\sqrt{\lambda(\lambda+1)}\\
	\pm2\sqrt{\lambda(\lambda+1)}&2\lambda+1
	\end{array}
	\right], 
\end{equation}
while $d=0$.
As is described in Sec.~\ref{secII}, $\lambda=\mathrm{sinh}^2r$ corresponds to the average photon number per mode. 

\subsubsection*{$\bold{Gaussian}$ $\bold{unitary}$ $\bold{operations}$}
Gaussian unitary operation is defined as an unitary operation transforming a Gaussian state to another Gaussian state, 
which includes the operations by linear optics and the second-order nonlinear process. 
Any Gaussian unitary operation acting on a Gaussian state is characterized by 
the following symplectic transformations: 
\begin{equation} 
\gamma\rightarrow S^T\gamma S,~~d\rightarrow S^Td,
\end{equation}
where $S$ is a symplectic matrix corresponding to the Gaussian unitary operation. 
The symplectic matrix for a beamsplitter on mode~A and mode~B is given by
\begin{equation}
S_{AB}^t=\left[
	\begin{array}{cccc}
	\sqrt{t}&\sqrt{1-t}&0&0\\
	-\sqrt{1-t}&\sqrt{t}&0&0\\
	0&0&\sqrt{t}&\sqrt{1-t}\\
	0&0&-\sqrt{1-t}&\sqrt{t}
	\end{array}
	\right], 
\label{sympbs}
\end{equation}
where $t$ is the transmittance of the beamsplitter. 
Hereafter, we simplify the description of a block diagonalized matrix like Eq.~(\ref{sympbs}) as 
\begin{equation}
S_{AB}^t=\left[
	\begin{array}{cc}
	\sqrt{t}&\sqrt{1-t}\\
	-\sqrt{1-t}&\sqrt{t}
	\end{array}
	\right]^{\oplus2}. 
\label{sympbsr}
\end{equation}

\subsubsection*{$\bold{Measurements}$}
We consider that the detectors D1, D2, D3 and D4 in Fig.~\ref{fig:BellSetup}(b) are on-off type, single photon detectors, 
namely, they only distinguish between vacuum and non-vacuum. 
Denoting the dark count probability of the detectors by $\nu$, 
the POVM elements of 
the on-off detectors are described by 
\begin{equation}
\hat{\Pi}^{\rm{off}}(\nu)=(1-\nu)\ketbra{0}{0}
\end{equation}
and
\begin{equation}
\hat{\Pi}^{\rm{on}}(\nu)=\hat{I}-\hat{\Pi}^{\rm{off}}(\nu),
\end{equation}
where $\hat{I}$ is the identity operator. 
The detection probability is calculated by introducing 
the characteristic functions of the POVM elements. 
Similar to the state, the characteristic function of the POVM element 
$\hat{\Pi}$ is given by $\chi_{\Pi}(\xi)=\mathrm{Tr}[\hat{\Pi}\hat{\mathcal{W}}(\xi)]$. 
When a single-mode Gaussian state $\hat{\rho}$ with characteristic function 
$\chi_\rho(\xi)=\mathrm{exp}(-\frac{1}{4}\xi^T\gamma\xi)$ is measured, 
the detection probability is given by 
\begin{eqnarray}
P_{\mathrm{on}}&=&\mathrm{Tr}[\hat{\rho}\hat{\Pi}^{\mathrm{on}}]=1-(1-\nu)\mathrm{Tr}[\hat{\rho}\ketbra{0}{0}]\nonumber\\
&=&1-\frac{2(1-\nu)}{\sqrt{\mathrm{det}(\gamma+I)}}. 
\end{eqnarray} 
\vspace{1mm}
\subsubsection*{$\bold{Linear}$~$\bold{loss}$}
The linear photon losses in the system such as 
a coupling efficiency to the single-mode fiber and 
imperfect quantum efficiency of the detectors are modeled 
by performing a beamsplitter transformation of
transmittance $t$ between the lossy mode and a vacuum mode, and 
tracing out the vacuum mode. 
The transformation of the linear loss on the state 
with covariance matrix $\gamma$ can be described as 
\begin{equation}
\mathcal{L}^t\gamma=K^T\gamma K+\alpha, 
\end{equation}
where $K=\sqrt{t}I$ and $\alpha=(1-t)I$. 

\subsection{Detection probabilities}
Using the above basic tools, 
we present the procedure to calculate the detection probabilities. 
As shown in Fig.~\ref{fig:BellSetup}(b) the entangled photon pair source consists of 
two TMSV sources over polarization modes. 
The covariance matrix of the output state is given by 
\begin{widetext}
\begin{equation}
\gamma_{{H_AV_AH_BV_B}}^{\mathrm{TMSV12}}=\left[
	\begin{array}{cccc}
	2\lambda_1+1&2\sqrt{\lambda_1(\lambda_1+1)}&0&0\\
	2\sqrt{\lambda_1(\lambda_1+1)}&2\lambda_1+1&0&0\\
	0&0&2\lambda_2+1&-2\sqrt{\lambda_2(\lambda_2+1)}\\
	0&0&-2\sqrt{\lambda_2(\lambda_2+1)}&2\lambda_2+1
	\end{array}
	\right]^{\oplus2}. 
\label{TMSV12}
\end{equation}
\end{widetext}
Note that the relative phase between TMSV1 and TMSV2 is set to $\pi$ as 
described in Eq.~(\ref{output3}).
In the experiment, the two TMSV sources are 
embedded in the Sagnac loop. In this case, the covariance matrix of the 
output state is transformed into~\cite{takeoka2015full} 

\begin{widetext}
\begin{equation}
\gamma_{{H_AV_AH_BV_B}}^{\mathrm{SL}}=\left[
	\begin{array}{cccc}
	2\lambda_1+1&0&0&2\sqrt{\lambda_1(\lambda_1+1)}\\
	0&2\lambda_2+1&-2\sqrt{\lambda_2(\lambda_2+1)}&0\\
	0&-2\sqrt{\lambda_2(\lambda_2+1)}&2\lambda_2+1&0\\
	2\sqrt{\lambda_1(\lambda_1+1)}&0&0&2\lambda_1+1
	\end{array}
	\right]^{\oplus2}. 
\label{TMSVSL}
\end{equation}
\end{widetext}

The covariance matrix in Eq.~(\ref{TMSVSL}) is first transformed by 
the (polarization-domain) beamsplitter operations $S^{\theta_A}_{H_AV_A}S^{\theta_B}_{H_BV_B}$.  
The covariance matrix after the transformation is given by 
 \begin{equation}
 \gamma_{{H_AV_AH_BV_B}}^{\mathrm{BS}}:=S^{\theta_B~T}_{H_BV_B}S^{\theta_A~T}_{H_AV_A}\gamma_{{H_AV_AH_BV_B}}^{\mathrm{SL}}S^{\theta_A}_{H_AV_A}S^{\theta_B}_{H_BV_B}.
 \end{equation}
The overall system losses including imperfect quantum efficiencies of the detectors are 
considered by performing linear-loss operations $\mathcal{L}^{\eta_1}_{H_A}$, $\mathcal{L}^{\eta_3}_{V_A}$, 
$\mathcal{L}^{\eta_2}_{H_B}$ and $\mathcal{L}^{\eta_4}_{V_B}$ on corresponding modes. 
The covariance matrix just before the detectors is given by 
\begin{widetext}
\begin{eqnarray}
\gamma_{{H_AV_AH_BV_B}}^{\mathrm{final}}&=&\mathcal{L}^{\eta_1}_{H_A}\mathcal{L}^{\eta_3}_{V_A}\mathcal{L}^{\eta_2}_{H_B}\mathcal{L}^{\eta_4}_{V_B} \gamma_{{H_AV_AH_BV_B}}^{\mathrm{BS}}\\
&=&K^{\eta_1\eta_3\eta_2\eta_4~T}_{H_AV_AH_BV_B} \gamma_{{H_AV_AH_BV_B}}^{\mathrm{BS}}K^{\eta_1\eta_3\eta_2\eta_4}_{H_AV_AH_BV_B}+\alpha^{\eta_1\eta_3\eta_2\eta_4}_{H_AV_AH_BV_B}
\end{eqnarray}
where
\begin{equation}
K^{\eta_1\eta_3\eta_2\eta_4}_{H_AV_AH_BV_B}=	
	\left[
	\begin{array}{cccc}
	\sqrt{\eta_1}&0&0&0\\
	0&\sqrt{\eta_{3}}&0&0\\
	0&0&\sqrt{\eta_2}&0\\
	0&0&0&\sqrt{\eta_4}
	\end{array}
	\right]^{\oplus2}. 
\end{equation}
and
\begin{equation}
\alpha^{\eta_1\eta_3\eta_2\eta_4}_{H_AV_AH_BV_B}=	
	\left[
	\begin{array}{cccc}
	1-\eta_1&0&0&0\\
	0&1-\eta_3&0&0\\
	0&0&1-\eta_2&0\\
	0&0&0&1-\eta_4
	\end{array}
	\right]^{\oplus2}. 
\end{equation}
\end{widetext}
The detection probabilities are calculated by performing $\hat{\Pi}^{\rm{on}/\rm{off}}(\nu)$ on corresponding modes. 
For example, the probability of observing clicks in D1 and D2 and no-clicks 
in D3 and D4 is given by 
\begin{widetext}
\begin{align}
&P(\mathrm{c1,c2,nc3,nc4}|\theta_{A},\theta_{B})=\mathrm{Tr}[\rho^{\gamma_{{H_AV_AH_BV_B}}^{\mathrm{final}}}\hat{\Pi}^{\rm{on}}_{H_A}(\nu)\hat{\Pi}^{\rm{on}}_{H_B}(\nu)\hat{\Pi}^{\rm{off}}_{V_A}(\nu)\hat{\Pi}^{\rm{off}}_{V_B}(\nu)]\\
&=\mathrm{Tr}[\rho^{\gamma_{{H_AV_AH_BV_B}}^{\mathrm{final}}}
(\hat{I}-(1-\nu)\ketbra{0}{0}_{H_A})(\hat{I}-(1-\nu)\ketbra{0}{0}_{H_B})(1-\nu)\ketbra{0}{0}_{V_A}(1-\nu)\ketbra{0}{0}_{V_B}]\\
&=\frac{4(1-\nu)^2}{\sqrt{\mathrm{det}(\gamma_{V_AV_B}^{\mathrm{final}}+I)}}-\frac{8(1-\nu)^3}{\sqrt{\mathrm{det}(\gamma_{H_AV_AV_B}^{\mathrm{final}}+I)}}\nonumber\\
&-\frac{8(1-\nu)^3}{\sqrt{\mathrm{det}(\gamma_{H_BV_AV_B}^{\mathrm{final}}+I)}}+\frac{16(1-\nu)^4}{\sqrt{\mathrm{det}(\gamma_{{H_AV_AH_BV_B}}^{\mathrm{final}}+I)}}.
\end{align}
\end{widetext}

\subsection{Calculation of $S$}
As in Eq.~(\ref{eq:CHSH}), $S$ is obtained by calculating 
$P(a=b|\theta_{A_i},\theta_{B_j})$ and $P(a\neq b|\theta_{A_i},\theta_{B_j})$ for $i, j=\{1,2\}$.  
For simplicity, omitting the conditions of the angles, 
these conditional probabilities are given by 
\begin{equation}
P(a=b)=P(+1,+1)+P(-1,-1)
\label{condpro1}
\end{equation}
and
\begin{equation}
P(a\neq b)=P(+1,-1)+P(-1,+1).
\label{condpro2}
\end{equation} 
In our model, each probability in the right hand side of Eq.~(\ref{condpro1}) and Eq.~(\ref{condpro2}) is calculated as follows: 

\begin{eqnarray}
P(-1,-1)&=&P(\mathrm{c1,c2,nc3,nc4}), \\
P(+1,-1)&=&P(\mathrm{c1,c2,c3,nc4})+P(\mathrm{nc1,c2,c3,nc4})\nonumber\\
&&+P(\mathrm{nc1,c2,nc3,nc4}),\\
P(-1,+1)&=&P(\mathrm{c1,c2,nc3,c4})+P(\mathrm{c1,nc2,nc3,c4})\nonumber\\
&&+P(\mathrm{c1,nc2,nc3,nc4}),\\
P(+1,+1)&=&1-P(-1,-1)\nonumber\\
&&-P(+1,-1)-P(-1,+1).
\end{eqnarray}


\begin{thebibliography}{10}

\bibitem{RevModPhys.79.135}
P.~Kok {\em et~al.},
\newblock Rev. Mod. Phys. {\bf 79}, 135 (2007).

\bibitem{RevModPhys.84.777}
J.-W. Pan {\em et~al.},
\newblock Rev. Mod. Phys. {\bf 84}, 777 (2012).

\bibitem{bell1964js}
J.~Bell,
\newblock Physics {\bf 1}, 195 (1964).

\bibitem{RevModPhys.86.419}
N.~Brunner, D.~Cavalcanti, S.~Pironio, V.~Scarani, and S.~Wehner,
\newblock Rev. Mod. Phys. {\bf 86}, 419 (2014).

\bibitem{PhysRevLett.98.230501}
A.~Ac\'{\i}n {\em et~al.},
\newblock Phys. Rev. Lett. {\bf 98}, 230501 (2007).

\bibitem{pironio2009device}
S.~Pironio {\em et~al.},
\newblock New Journal of Physics {\bf 11}, 045021 (2009).

\bibitem{colbeck2012free}
R.~Colbeck and R.~Renner,
\newblock Nature Physics {\bf 8}, 450 (2012).

\bibitem{giustina2013bell}
M.~Giustina {\em et~al.},
\newblock Nature {\bf 497}, 227 (2013).

\bibitem{PhysRevLett.111.130406}
B.~G. Christensen {\em et~al.},
\newblock Phys. Rev. Lett. {\bf 111}, 130406 (2013).

\bibitem{PhysRevLett.115.250401}
M.~Giustina {\em et~al.},
\newblock Phys. Rev. Lett. {\bf 115}, 250401 (2015).

\bibitem{PhysRevLett.115.250402}
L.~K. Shalm {\em et~al.},
\newblock Phys. Rev. Lett. {\bf 115}, 250402 (2015).

\bibitem{PhysRevLett.23.880}
J.~F. Clauser, M.~A. Horne, A.~Shimony, and R.~A. Holt,
\newblock Phys. Rev. Lett. {\bf 23}, 880 (1969).

\bibitem{PhysRevLett.120.010503}
Y.~Liu {\em et~al.},
\newblock Phys. Rev. Lett. {\bf 120}, 010503 (2018).

\bibitem{Lijiong2018}
L.~Shen {\em et~al.},
\newblock ArXiv e-prints  (2018), arXiv:1805.02828.

\bibitem{PhysRevA.91.012107}
V.~Caprara~Vivoli {\em et~al.},
\newblock Phys. Rev. A {\bf 91}, 012107 (2015).

\bibitem{takeoka2015full}
M.~Takeoka, R.-B. Jin, and M.~Sasaki,
\newblock New Journal of Physics {\bf 17}, 043030 (2015).

\bibitem{PhysRevA.93.042328}
K.~P. Seshadreesan, M.~Takeoka, and M.~Sasaki,
\newblock Phys. Rev. A {\bf 93}, 042328 (2016).

\bibitem{cirel1980bs}
B.~Cirelson,
\newblock Lett. Math. Phys. {\bf 4}, 93 (1980).

\bibitem{PhysRevA.47.R747}
P.~H. Eberhard,
\newblock Phys. Rev. A {\bf 47}, R747 (1993).

\bibitem{Jin:14}
R.-B. Jin {\em et~al.},
\newblock Opt. Express {\bf 22}, 11498 (2014).

\bibitem{Miki:17}
S.~Miki, M.~Yabuno, T.~Yamashita, and H.~Terai,
\newblock Opt. Express {\bf 25}, 6796 (2017).

\bibitem{klyshko1980use}
D.~Klyshko,
\newblock Sov. J. Quantum Electronics {\bf 10}, 1112 (1980).

\end{thebibliography}

\end{document}